\RequirePackage{fix-cm}
\documentclass{svjour3}                     
\smartqed  
\usepackage{amssymb}
\usepackage{graphics}
\usepackage{amsmath}
\usepackage{lineno,hyperref}
\modulolinenumbers[5]
\usepackage{graphicx} 
\usepackage{bm}
\usepackage{mathtools,slashed}
\usepackage{appendix}
\journalname{Foundations of Physics}
\begin{document}

\title{Polarization of vacuum fluctuations: source of the vacuum permittivity and speed of light
}

\titlerunning{vacuum fluctuations: source of the vacuum permittivity and speed of light}        

\author{G. B. Mainland        \and
        Bernard Mulligan 
}


\institute{G. B. Mainland \at
              Department of Physics, The Ohio State University at Newark,  1179 University Dr., Newark, OH 43055-1797, USA\\
              Tel.: 1 (740) 616-4632\\
              Fax: 1 (603) 319-4579\\
              \email{mainland.1@osu.edu}           
           \and
           Bernard Mulligan \at
              Department of Physics, The Ohio State University, Columbus, OH 43210, USA\\
              Tel.: 1 (917) 969-3710\\
              Fax: 1 (212) 712-0849\\
              \email{mulligan.3@osu.edu} 
}

\date{Received: 24 August 2019 / Accepted: 7 March 2020}

\maketitle

\begin{abstract}
There are two types of fluctuations in the quantum vacuum: type 1 vacuum fluctuations are on shell and can interact with matter in specific, limited ways that have observable consequences; type 2 vacuum fluctuations are off shell and cannot interact with matter.  A photon will polarize a type 1, bound, charged lepton-antilepton vacuum fluctuation in much the same manner that it would polarize  a dielectric, suggesting the method used here for calculating the permittivity  $\epsilon_0$ of the  vacuum.  In a model that retains only leading terms, $\epsilon_0 \cong   (6\mu_0/\pi)(8e^2/\hbar)^2= 9.10\times 10^{-12}$ C/(Vm).  The calculated value for $\epsilon_0$ is  2.7\% more than the accepted value.  The permittivity of the vacuum, in turn, determines the speed $c$ of light in the vacuum. Since the vacuum is at rest with respect to every inertial frame of reference, $c$ is the same in every inertial reference frame.
\end{abstract}

\section{Introduction:  the vacuum as a dielectric}
\label{sec:1}

The idea that  vacuum fluctuations\footnote{Type 1 vacuum fluctuations, which have observable effects,  and type 2 vacuum fluctuations, which have no observable effects, will be discussed in detail in Sec. \ref{sec:2}.}  play a role in determining the permittivity of the vacuum is very old:  in 1934 Furry and Oppenheimer\cite{Furry:34} wrote that vacuum fluctuations of charged particle-antiparticle pairs would affect the value of the dielectric constant of the vacuum: ``Because of the polarizability of the nascent pairs, the dielectric constant of space into which no matter has been introduced differs from that of truly empty space.''  Also in 1934 the idea of treating  the vacuum as a medium with electric and magnetic polarizability was discussed by  Pauli and Weisskopf \cite{Pauli:94} and again two years later by Weisskopf\cite{Weisskopf:94}. In 1957 Dicke\cite{Dicke:57} wrote about the possibility that the vacuum could be considered as  a dielectric medium. Wilczek's  2008  book {\it The Lightness of Being}\cite{Wilczek:08} expressed the fundamental characteristics of space and time as properties of the Grid, ``the entity we perceive as empty space.  Our deepest physical theories  reveal it to be highly structured; indeed, it appears as the primary ingredient of  reality.''  Recently the possibility that the properties of the vacuum determine, in the vacuum, the speed of light and the permittivity have been explored by a number of authors\cite{Leuchs:10,Leuchs:13,Urban:13,Mainland:19}. 

A basic postulate of physics is that the properties of a physical system are determined by the structure of that system.  The only events that occur in the quantum vacuum are  the spontaneous appearance and subsequent disappearance of  vacuum fluctuations that occur for every type of particle-antiparticle pair, including pairs for which the particle is its own antiparticle, as is the case for photons. 

The photon has spin 1; a vacuum fluctuation of the electromagnetic field must have spin 0.\footnote{Planck was unaware of photon spin, and his original derivations were independent of that spin. For a thorough discussion of Planck's derivation consistent with photon spin, see \cite{Itzykson:05}.} It is important to point out that, other than the preceding comment, this article does not discuss fluctuations of the QED vacuum, on which there already exists considerable literature\cite{Milonni:94,Leonhardt:97,Meis:17a} and references therein. Accordingly, here it will not be possible to address questions as to whether the cosmological constant is affected by or results from QED fluctuations.  In this regard, Leonhardt\cite{Leonhardt:19} has shown that the electromagnetic contribution to the cosmological constant need not include fluctuations of the QED quantum vacuum. It will also not be possible to address the Casimir force, long thought to be the result of zero-point energy of the  QED vacuum.  However, the energy of the QED vacuum is no longer considered to be responsible for the Casimir force\cite{Grundler:13,Schwinger:78,Milonni:82}. A recent article by Meis\cite{Meis:18} provides insights into understanding the electromagnetic quantum vacuum and the possibility of eliminating infinities associated with it.

Only fluctuations of the quantum fermion vacuum will be considered here, and a (real) test photon interacting with those  fluctuations will be shown to determine the values of $\epsilon_0$ and $c$.  To minimize the violation of conservation of energy allowed by the Heisenberg uncertainty principle and to avoid violating conservation of angular momentum, in any inertial frame a vacuum fluctuation consisting of a charged fermion-antifermion pair must appear with its center of mass at rest and in the least energetic bound state that has zero angular momentum.  In addition to the above theoretical reason, there is a second, physical reason for expecting charged  fermion-antifermion vacuum fluctuations to appear as bound states: the value of the permittivity $\epsilon$  of a physical dielectric is determined by the electromagnetic properties of the physical dielectric and results from atoms or molecules in the dielectric being polarized by an electric field (or photons). Just as an effective spring constant is associated with the resultant relative motion of the constituents of the atoms or molecules, an effective spring constant exists for a bound, charged fermion-antifermion pair.  Transient fermion-antifermion atoms \textendash \, parapositronium for an electron-positron pair \textendash \, are of primary interest in this article, and, as will be demonstrated,  these transient atoms can be observed indirectly by the effect of their interaction with photons moving through the vacuum. 

If the value of the permittivity of the vacuum $\epsilon_0$ is determined by the structure of the vacuum, it should be possible to calculate $\epsilon_0$  from that structure by examining the (polarizing) interaction of  photons ``introduced'' into the vacuum as test particles.   The model of the vacuum discussed here yields an approximate formula for the permittivity of the vacuum\footnote{SI units are used throughout this article except in the decay rate calculation in the ``Appendix'' where, following convention, $\hbar$ and $c$ are replaced by 1. After reinstating factors of $\hbar$ and $c$, from \eqref{eqn:57} the electromagnetic decay rate for a photon-excited, parapositronium VF  is $\Gamma_{p-P_s} \, = (\alpha^5 m_e c^2)/\hbar$.  The mean lifetime of the state is $1/\Gamma_{p-P_s} \, \cong \, 6.2 \times 10^{-11}s$.}
\begin{equation}\label{eqn:1}
\epsilon_0 \cong   \frac{ 6\mu_0}{\pi}\left(\frac{8e^2}{\hbar}\right)^2= 9.10\times 10^{-12}\rm \frac{C}{Vm}\, .
\end{equation}
In the above equation $e$ is the magnitude of the (renormalized) charge of an electron,  $\hbar$ is Planck's constant divided by $2\pi$, and $\mu_0$\footnote{The value of $\mu_0$ was originally arbitrarily chosen so that the rationalized meter-kilogram-second unit of current was equal in size to the ampere in the ``electromagnetic (emu)'' system.  As a consequence,  $\mu_0$, which is a measurement-system constant, not a property of the vacuum, was defined to be exactly $\mu_0\equiv 4\pi \times 10^{-7}$ H/m.  Since physicists arbitrarily chose the value of $\mu_0$,  $\mu_0$ could neither be measured experimentally nor calculated theoretically.   As of May 20, 2019,  $\mu_0$ is no longer the defined constant $4\pi \times 10^{-7}$ H/m. Instead  $\mu_0$ is defined by the equation $\mu_0\equiv2\alpha h/(ce^2)$, which follows from the definition of the fine-structure constant and $c=1/\sqrt{\mu_0\epsilon_0}$. Similarly, $\epsilon_0 \equiv e^2/(2 \alpha h c)$, which follows from the definition of the fine-structure constant.  These new definitions for $\mu_0$ and $\epsilon_0$ satisfy the condition $\mu_0\epsilon_0=1/c^2$, as they must.  The quantities $c, h$ and $e$ are now defined to have specific values, and $\alpha$ is determined experimentally. The changes on May 20, 2019, only impact the value of $\mu_0$ in about the tenth significant figure and beyond.  [See Bureau International des Poids et Mesures (BIPM): website (https://physics.nist.gov/cuu/Units/)].  The permeability $\mu$ of a magnetic medium can, of course, be determined experimentally and calculated theoretically. In Sec. 6  the effect of vacuum fluctuations on the permeability of the vacuum is calculated.  There is no effect so the value of $\mu_0$ is unaffected by fluctuations of the charged fermion-antifermion vacuum.} is the permeability of the vacuum.  The calculated value for $\epsilon_0$ is  2.7\% more than the accepted value. The calculation of $\epsilon_0$  has been  simplified, and the numerical accuracy has been reduced, by including  (a) only contributions to lowest order in what turns out to be an expansion in powers of the fine-structure constant $\alpha$ and (b) only the interactions of photons with bound states of charged lepton-antilepton vacuum fluctuations.

Once $\epsilon_0$ has been calculated, a formula for the speed $c$ of light in the vacuum is obtained from  $c=1/\sqrt{\mu_0\epsilon_0}$ and the formula for $\epsilon_0$ in \eqref{eqn:1},
\begin{equation}\label{eqn:2}
c \cong   \sqrt{\frac{\pi}{6}}\frac{\hbar}{8e^2\mu_0}= 2.96\times 10^8\rm{m/s}\,.
\end{equation}
The value of $c$ calculated from this model is 1.4\% less than the defined value $c=3.00 \times 10^8$m/s.  Thus the calculation in the following sections shows that a photon in the vacuum is slowed by interacting with and exerting a polarizing effect on a bound, charged lepton-antilepton vacuum fluctuation. 

An observer in any inertial frame is unable to detect the relative motion of the inertial frame and the quantum vacuum. Therefore the observer  would conclude that the inertial frame and the vacuum are at rest with respect to each other. As Leonhardt et al.\cite{Leonhardt:18} state,``In free space the vacuum is Lorentz invariant,  so a uniformly moving observer would not see any effect due to motion, but an accelerated observer would.  The latter is known as the Unruh effect\cite{Unruh:76}.''  The interaction of photons with fermion vacuum fluctuations determines the speed of light in the vacuum, and in the vacuum fermion vacuum fluctuations  are at rest with respect to any inertial frame. It  then follows that $c$  is the same in any inertial frame, which is one of two postulates in the 1905 paper ``On the Electrodynamics of Moving Bodies''\cite{Einstein:05} in which  special relativity was introduced by Einstein. The derivation here of the value of $c$  provides a theoretical explanation for why Einstein's postulate is true, obviating the need for the postulate as a separate concept.

There is a second theoretical approach for calculating $c$ that would satisfy Einstein's  postulate that $c$  is the same in any inertial frame:  if $c$ could be calculated from an internal consistency condition within QED,  then $c$ would be the same in every inertial frame since the equations of physics are the same in every inertial frame. Such a calculation has never been completed.

\section{Vacuum fluctuations}
\label{sec:2}

Vacuum fluctuations in the quantum fermion vacuum refer to two distinct entities, labeled here as type 1 and type 2, with very different properties.   The field-theoretical proofs that vacuum fluctuations exist allow the two types to be clearly distinguished. Based on those proofs, a type 1 vacuum fluctuation is defined to be a vacuum fluctuation (net charge zero) that  appears as an external particle in a Feynman diagram, is on shell, and exists only for a time allowed by the Heisenberg uncertainty principle\cite{Bjorken:65,Thirring:58}.  A type 1 vacuum fluctuation can consist of a charged or uncharged particle-antiparticle pair. 

 The appearance of a type 1 vacuum fluctuation is a stochastic process: as such, either a type 1 vacuum fluctuation does not appear at all or it appears on shell. Once charged lepton-antilepton type 1 vacuum fluctuations spontaneously appear, as will be discussed in Sec. \ref{sec:3}  they can interact with physical particles, but only in very limited ways. And if they do not interact, the charged pair will simply annihilate, returning the borrowed energy to the vacuum. The calculation here relies on the ``introduction'' of a (real) photon into the quantum vacuum as a test particle. The interaction of that photon with a  bound, charged lepton-antilepton type 1 vacuum fluctuation  creates a polarized  type 1 vacuum fluctuation that subsequently decays, emitting a real photon. The  polarized, type 1 vacuum fluctuation is a quasi-stationary state,  a state for which the lifetime of the state is large in comparison with the uncertainty in the lifetime\cite{Davydov:76}. In addition to interacting with a photon to form a quasi-stationary state, a type 1 vacuum fluctuation can also interact through the annihilation of a particle and antiparticle.  For example, a real electron can annihilate with the positron of a  type 1 vacuum fluctuation.  The electron that was part of the  type 1 vacuum fluctuation then becomes a real electron with a location different from that of the original, physical electron,  giving rise to zitterbewegung (``trembling motion'')\cite{Thirring:58}.  

Texts on field theory define type 2 vacuum fluctuations to be vacuum diagrams \cite{Jauch:76}, which are the same as vacuum bubbles\cite{Bjorken:65}, a class of perturbation effects in quantum field theories with Feynman diagrams that look like bubbles that originate from and terminate in the vacuum.  Vacuum diagrams do not make a contribution to physical processes\cite{Bjorken:65,Peskin:95,Jauch:76}; consequently,  type 2 vacuum fluctuations do not have observable effects of the type described here.

Vacuum fluctuations are sometimes confused with virtual particles, for which a widely accepted definition exists. For example Penrose\cite{Penrose:05} defines a virtual particle as follows:   A virtual particle is ``off-shell'' and ``can occur only in the interior of a Feynman graph.''   A virtual particle-antiparticle pair annihilates within a time that is ``short enough  that the energy required to produce the pair falls within the uncertainties of Heisenberg's time-energy relation$\dots$.'' Both the definition of a virtual particle by Langacker\cite{Langacker:17} and the use of the term ``virtual particle'' by Berestetskii et al.\cite{Berestetskii:82}  and by Gottfried and Weisskopf  \cite{Gottfried:86} are consistent with Penrose's definition. 

Virtual particles,  real particles,  type 1 vacuum fluctuations and type 2 vacuum fluctuations all have different properties:  virtual particles are off shell and transitory, real particles are on shell  and their energy is permanently present, type 1 vacuum fluctuations are on shell and transitory and type 2 vacuum fluctuations are a collection of interacting, virtual particles that arise in and then disappear back into the vacuum so they are off shell and transitory.   Perhaps because virtual particles and both types of vacuum fluctuations are transitory,  the two terms ``virtual particles'' and ``vacuum fluctuations'' are sometimes used as synonyms. They are not. In this article the acronym VF refers exclusively to a type 1 vacuum fluctuation that consists of a charged particle bound with angular momentum zero to its antiparticle.

Field theory provides an explanation for the source of the energy available for the creation of a VF and a proof that VFs must exist.  The structure of an atom that is a VF does not play a significant role in the discussion  of the production of that VF.  (Of course, the structure of the VF, which results from the electromagnetic interaction of the charged lepton-antilepton pair, is important for  the calculation of the decay rate of the atom that is a VF.)  Thus an atom consisting of a charged lepton and corresponding antilepton in its  ground state with zero angular momentum can, as far as its creation is concerned, be approximately represented  by a free, neutral,  spin-0, Klein-Gordon field $\phi(x)$, as first suggested by Pauli and Weisskopf\cite{Pauli:94} and elaborated on by Wentzel\cite{Wentzel:03}.  Using a field to describe a particle with internal degrees of freedom is discussed in Ref.\cite{Thirring:58}; representing a particle by a quantum field is essential to understanding the mathematical structure of a VF of a quantum field. Indeed, it is the quanta of a free field that behave as free particles. 

The vacuum expectation value of the free Hamiltonian $H$ for a neutral, spin-0 field representing an atom with mass $M$  and momentum $\mathbf{p}=\hbar\mathbf{k}$ can be written in the form,
\begin{equation}\label{eqn:101}
(0|H|0)= \frac{1}{2}\sum_\mathbf{k} \hbar \omega_\mathbf{k}\,,\hspace{0.3 cm}\omega_\mathbf{k}=+\sqrt{c^2\mathbf{k}^2+\frac{M^2c^4}{\hbar^2}} \,.
\end{equation}
For a lepton-antilepton VF the mass $M$ in the above equation is twice the mass of the lepton minus the binding energy of the lepton-antilepton bound state. The energy in the vacuum, called the zero-point energy, appears on the right-hand side of \eqref{eqn:101}. The sum over $k$ is infinite because there are an infinite number of cells in $\mathbf{k}$-space, implying that the energy in the vacuum is infinite.  To ensure Lorentz invariance and translation invariance of the vacuum, VFs must be created at rest in any inertial frame, implying that $\mathbf{k} =0$. It then follows that the only zero-point energy available to create a VF at rest is the single, finite term $\hbar\omega_{\mathbf{k}=0}=Mc^2$. The factor of  1/2 in \eqref{eqn:101} occurs because  VFs  are, on average, present only for half of the time.
  
To show that VFs  of charged lepton-antilepton pairs must exist,  note that  the expectation value  of the product of the free field $\phi(x)$ at two different locations $x$ and $x^\prime$ is\cite{Thirring:58,Bjorken:65}
\begin{equation}\label{eqn:105}
(0|\phi(x)\phi(x^\prime)|0)=\int_{k_0=\omega_\mathbf{k}}\hspace{-0.2 cm}\hbar \,  \frac{{\rm d}^3k}{(2\pi)^32\omega_\mathbf{k}}e^{-ik^\mu(x_\mu-x^\prime_\mu)}\,.\end{equation}
Eq. \eqref{eqn:105} has the feature that $(0| \phi^2(x) |0)$ is infinite.   However any vacuum fluctuation must have a finite size as will become  apparent in the next section. As a consequence, the symbol   $(0| \phi(x) \phi (x') |0)$ has no physical meaning unless $x$ and  $x^\prime$ are averaged over the size associated with the vacuum fluctuation. This point is particularly emphasized by\cite{Bjorken:65}.  The fact that the vacuum expectation value of the product $\phi(x)\phi(x^\prime)$ is nonzero demonstrates that the  free field $\phi(x)$ in the vacuum is nonzero.  
 
 Because VFs occur for noninteracting fields, the quanta associated with the fields are on shell: note from \eqref{eqn:105} that the integral over ${{\rm d}^3k}$ is subject to the constraint  $k_0=\omega_\mathbf{k}$.  It then immediately follows that $E^2-(\mathbf{p}c)^2= (\hbar \omega_\mathbf{k})^2-(\hbar \mathbf{k}c)^2=M^2\,c^4$, which is the on-shell condition. 
 
VFs only affect properties of the vacuum and do not alter existing predictions of QED\cite{Leuchs:20}.  If a photon were emitted by some process, it could interact with a VF.   But when the VF annihilated, a photon identical to the incident photon would be emitted. The photon emitted by the original process would just be traveling through the vacuum as usual.   Also, a photon emitted when a VF annihilated could interact with matter.  This would just be the process of an external photon being incident on and interacting with matter.  Thus, the existence of  fluctuations of charged fermion fields does not affect QED calculations.  However, if a VF interacted in a manner that supplied energy equal to or greater than the energy originally borrowed from the vacuum for its creation, the VF could become a physical particle-antiparticle pair, and the pair would interact as would any other physical particles. 

\section{Calculation of dielectric properties of the vacuum: the characteristics of a VF} 
\label{sec:3}

An electron-positron VF appears in the vacuum at rest as parapositronium\cite{Deutsch:52},  a singlet spin state with the lowest positronium bound-state energy and zero angular momentum.  Here attention is restricted to parapositronium since the corresponding results for muon-antimuon and tau-antitau VFs immediately follow by replacing the electron mass with the muon or tau mass, respectively.  The  non-relativistic  binding energy for parapositronium, $E_{p-Ps}$,  is  obtained from the $n = 1$ binding energy of hydrogen\cite{Schiff:55} by changing the reduced mass: $\mu_{\rm{hydrogen}}\simeq m_e\rightarrow \mu_{p-Ps}=m_e/2$, where $m_e$ is the mass of an electron:
\begin{equation}\label{eqn:3}
E_{p-Ps} = -\frac{ (m_e/2)e^4 }{2(4\pi \epsilon_0)^2 \hbar^2} =-\frac{m_e\alpha^2c^2}{4} \,. 
\end{equation}
In \eqref{eqn:3} $\alpha=e^2/(4 \pi \epsilon_0 \hbar c)$  is the fine-structure constant. 

From both the classical\cite{Kramers:24,Rossi:59,Feynman:63} and quantum\cite{Sokolov:84,Holstein:92} calculations of the permittivity of  physical matter consisting of atoms (or molecules),  it follows that for matter to possess permittivity,  the atoms must be able to oscillate when interacting with an electric field (or photons).  Taking the $x$-axis to point in the direction of the electric field, since the atom oscillates along the direction of the electric field, only the oscillatory properties of the atom  along the $x$-axis are of significance in the interaction.  Thus to first order in $\alpha$, when calculating permittivity, the interaction of the electromagnetic field with the atom can be described by a one-dimensional oscillator.  Although parapositronium VFs do not exhibit resonant behavior, the potential that binds the electron and positron into the parapositronium VF has a relative minimum and, for small oscillations, acts as if it were a harmonic oscillator with a resonant frequency.

As pointed out by Feynman\cite{Feynman:64}, when interacting  with an electric field,  an atom in its ground state interacts with the electric field as if it were a harmonic oscillator with the first two  energy levels separated by the binding energy of the atom. Since adjacent harmonic oscillator energy levels are separated by an energy $\hbar \omega^0$, from  \eqref{eqn:3},
\begin{equation}\label{eqn:4}
\omega^0=\frac{|E_{p-Ps}|}{\hbar}=\frac{m_e\alpha^2c^2}{4\hbar} \,.
\end{equation}

The effective spring constant of parapositronium  is $K_{p-Ps}=\mu_{p-Ps} (\omega^0)^2$ so the Hamiltonian $H^0$ describing an oscillating parapositronium VF is
\begin{equation}\label{eqn:5}
H^0=\frac{-\hbar^2}{2\mu_{p-Ps}}\frac{{\rm d^2}}{{\rm d}x^2}+ \frac{1}{2}\mu_{p-Ps} (\omega^0)^2 x^2\,.
\end{equation}
Using the formula for the energy\cite{Schiff:55}  of a harmonic oscillator in one dimension, the energy eigenvalues $E_n^0$ of the above Hamiltonian are  
\begin{equation}\label{eqn:6}
E_n^0=\hbar \omega^0(n+\frac{1}{2})\,;\; \;n=0,1,2\dots\,.
\end{equation}

For parapositronium that is a VF, the Heisenberg uncertainty principle is 
\begin{equation}\label{eqn:7}
\Delta E_{\rm p-Ps}\, \Delta t_{\rm p-Ps}\geq\frac{\hbar} {2} \, .
\end{equation}
Denoting the mass of an electron (or positron) by $m_e$,  $\Delta E_{\rm p-Ps}$ is the energy $2m_ec^2$ for the production  of parapositronium that is a VF\footnote{The binding energy of parapositronium, which is small in comparison with  $2m_ec^2$, is being neglected.}. The minimum time $\Delta t$ is the average lifetime $\Delta t_{\rm p-Ps}$  for the existence  of parapositronium that is a VF.  Thus \eqref{eqn:7} yields 
\begin{equation}\label{eqn:8}
\Delta t_{\rm p-Ps}= \frac{\hbar}{4m_ec^2} \, .
\end{equation}
During the time $\Delta t_{\rm p-Ps}$, a beam of light  travels a distance $L_{\rm p-Ps}$ given by
\begin{equation}\label{eqn:9}
L_{\rm p-Ps}= c \,\Delta t_{\rm p-Ps}= \frac{\hbar}{4m_ec} \, .
\end{equation}

Since a parapositronium VF appears from the vacuum at essentially a single location and since nothing can travel faster than the speed of light, while they exist the maximum distance between the electron and positron in parapositronium  is  $L_{\rm p-Ps}$.  Having already borrowed energy $2m_ec^2$ from a volume $\mathbb{V_{\rm p-Ps}}\equiv (L_{\rm p-Ps})^3$ of the vacuum, a second parapositronium VF is unlikely to form  in the same volume while the first still exists.  This suggests the ansatz, 
\begin{equation}\label{eqn:900}
\mbox{Number density of parapositronium VFs}= \frac{1}{(L_{\rm p-Ps})^3}\equiv\frac{1}{\mathbb{V}_{\rm p-Ps}} \, ,
\end{equation}
a result that can  immediately be generalized to other charged lepton-antilepton VFs and quark-antiquark VFs. 

The center of mass of a VF will remain fixed, accompanied by an internal zitterbewegung of the VF, giving the VF its size\cite{Gersch:92,Corinaldesi:63}.  The zitterbewegung internal to the  VF has an amplitude\cite{Barut:81}\footnote{Reference\cite{Sakurai:67} is referring to zitterbegung for spin-1/2 particles, as originally discussed by Schr\"odinger. It has long been recognized the zitterbewegung is also associated with Klein-Gordon (spin-0) particles \cite{Corinaldesi:63}. This result is also reflected  in the paper by Gersch\cite{Gersch:92}.}
\begin{equation}\label{eqn:901}
\frac{\hbar}{4m_ec} \equiv L^{\rm Z}_{\rm p-Ps} = L_{\rm p-Ps}\,,
\end{equation}
where the final equality follows from \eqref{eqn:9}. Eq. \eqref{eqn:901} allows a length $L^{\rm Z}_{\rm p-Ps}$ to be associated with the zitterbewegung.  The volume $(L^{\rm Z}_{\rm p-Ps})^3=(L_{\rm p-Ps})^3 \equiv \mathbb{V}_{\rm p-Ps}$ results from zitterbewegung and represents the volume of a VF. Requiring that there be only one VF in the volume of a VF as calculated from zitterbewegung yields the same formula previously obtained for the number density of VFs.  As a result of zitterbewegung, the VF has a size $\mathbb{V}_{\rm p-Ps}$ over which it must be averaged, thus removing the infinity in $(0|\phi^2(x)|0)$ discussed following \eqref{eqn:105}.

The number of lepton-antilepton VFs  per unit volume has values from $1.12 \times 10^{39}$/m$^3$ for electron-positron VFs to  $4.70 \times 10^{49}$/m$^3$ for tau-antitau VFs.  In contrast the number density of atoms or molecules of an ideal gas at STP is  $2.68 \times 10^{25}$/m$^3$. It is possible for the number density of VFs to be many orders of magnitude greater than the number density of atoms or molecules of an ideal gas because a VF cannot exert a force. 

First consider the electromagnetic force: if a VF has not already absorbed radiation, it cannot spontaneously emit radiation. If it did, the radiated photons would exist after the VF has disappeared back into the vacuum, permanently violating conservation of energy.  If a VF has interacted with a photon, when the VF vanishes back into the vacuum, the VF must emit a photon identical to the incident photon in order to conserve energy, momentum, and angular momentum. Since a VF cannot ``permanently'' exchange a photon with either another VF or a physical quantum, it cannot exert a force on either.   Similar arguments verify that VFs cannot exert a force of any type. There is also a simple, classical argument that a VF cannot exert a force on a physical particle:  if it could, the energy associated with the work done by the force would remain after the VF vanished back into the vacuum, permanently violating conservation of energy. 

The dipole moment (operator) $p_x$ of an atom in the presence an electromagnetic wave with its electric field along the $x$-axis is  $p_x=e(x_+-x_- )\equiv ex$, where $x_-$ and $x_+$ are, respectively, the coordinates of the electron and positron.  The Hamiltonian $H^1$ describing the interaction of the electric dipole of the atom with the electromagnetic wave $E_x=E_0\cos\omega t$ is
\begin{equation}\label{eqn:10}
H^1=-\mathbf{p}\cdot\mathbf{E(t)}=-exE_0\cos \omega t\,.
\end{equation}

When examining \eqref{eqn:10}, the following question immediately arises: when a VF interacts with a photon, does it typically interact with one or multiple photons? This is the essential question of photon-VF interactions; the answer is that if a lepton-antilepton VF interacts with a photon at all, it essentially always interacts with only one photon.  As a result the electric field of each individual photon polarizes each individual VF with which it interacts, as suggested by the title of the paper. 

To see this consider the photon number density $N_\gamma$ of a laser beam with cross-sectional area $A$ and wavelength $\lambda$. During a time $\Delta t$ the laser emits  energy $\Delta E$, which equals  the energy per unit volume $N_\gamma hc/\lambda$ of the beam multiplied by the volume $Ac\Delta t$ of the pulse,
\begin{equation}\label{eqn:11}
\Delta E=\frac{N_\gamma hc}{\lambda}(Ac\Delta t)\,.
\end{equation}
Solving \eqref{eqn:11} for $N_\gamma$ and noting that the power $P=\Delta E/\Delta t$,
\begin{equation}\label{eqn:12}
N_\gamma =\frac{P \lambda}{hc^2A}\,.
\end{equation}
In a 6,000 W, CO$_2$ cutting laser with  a wavelength of 10$\mu$m and a beam radius of 0.16 mm, the number density $N_\gamma$ of photons is on the order of $10^{22}$ photons/m$^3$.  Even in such an intense laser beam, the number density of lepton-antilepton VFs is  is much  greater than the number density of photons.  

\section{Calculation of dielectric properties of the vacuum: dipole moment of VFs} 
\label{sec:4}

When a photon interacts with a parapositroniun VF, for example, and combines its energy, momentum and angular momentum  with that of the parapositroniun VF, the electric field of the photon that interacts with the parapositronium VF is the electric field at the instant $t_i$ that the interaction takes place:  $E(t_i) = E_0\cos \omega t_i \equiv \mathbb{E}_0$.  Thus the electric field $\mathbb{E}_0\,\mathbf{\hat{x}}$ describes the interaction of a single photon with a VF and  \eqref{eqn:10} must be replaced by
\begin{equation}\label{eqn:13}
H^{1 ^{V\!F}}=-exE_0\cos \omega t_i \equiv -ex\mathbb{E}_0\,.
\end{equation}

The polarization of a bound, lepton-antilepton  VF (a parapositronium vacuum fluctuation, for example) can be obtained from a quantum mechanics calculation that is similar to the calculation of the polarization of an ordinary dielectric. There are, however,  two major differences: (1) At the instant a photon interacts with a vacuum fluctuation, the value of the electric field associated with the photon is the value at the moment of interaction. (2) Since an atom or molecule in a dielectric usually has an axis of orientation,  a factor of one third is required when averaging over the three possible axes of orientation of the atom or molecule\cite{Cook:75}. Because vacuum fluctuations are spherically symmetric and thus do not have an orientation axis, the factor of one third is not present.

The polarization that  results when a  parapositronium atom in its ground state  interacts with a photon is readily calculated.  Let $\psi_n^0(x)$ be solutions to the time-independent, one-dimensional, unperturbed Schr\"odinger equation,
\begin{equation}\label{eqn:14}
H^0\psi_n^0(x)=E_n^0\psi_n^0(x)\,,
\end{equation}
where   $H^0$ and $E_n^0$ are given by \eqref{eqn:5} and \eqref{eqn:6}, respectively.   From stationary perturbation theory, to lowest order in the perturbation, the ground-state solution $\psi^1_0(x)$ of the Schr\"odinger equation corresponding to the Hamiltonian $H^0+H^{1^{V\!F}}$ is\cite{Schiff:55}
\begin{equation}\label{eqn:15}
\psi^1_0(x)=\,\psi^0_0(x)+\!\sum_{n^\prime\neq 0}\frac{\psi^0_{n^\prime}(x)}{E^0_0-E^0_{n^\prime}}\int_{-\infty}^{\infty}\!\!\rm{d}x^\prime\,\psi^{0}_{n^\prime}(x^\prime)^*(-ex^\prime\mathbb{E}_0)\psi^0_0(x^\prime)\,.
\end{equation}
 The only nonzero contribution to the sum occurs for $n^\prime=1$.  Then in the above equation  the  integral equals  $-e\mathbb{E}_0\sqrt{\hbar/(2\mu\omega^0)}$, and the energy difference  $E^0_0-E^0_1=-\hbar\omega^0$. Thus \eqref{eqn:15} becomes
\begin{equation}\label{eqn:16}
\psi^1_0(x)=\,\psi^0_0(x)+\frac{e\mathbb{E}_0}{\sqrt{2 \mu \hbar(\omega^0)^3}}\psi^0_1(x)\,.
\end{equation}
The expectation value $\langle p^{V\!F}\!\rangle$ of the electric dipole moment in the state characterized by $\psi^1_0(x)$ is 
\begin{equation}\label{eqn:17}
\langle p^{V\!F}\!\rangle=\int_{-\infty}^\infty{\rm d}x\,\psi^{1}_0(x)^*\,ex\,\psi^1_0(x)=\frac{(e^2/\mu)\mathbb{E}_0} {(\omega^0)^2}\,. 
\end{equation}

VFs are not restricted to electron-positron pairs.  To allow for  more than one type of charged particle-antiparticle VF that binds into an atom, an index $j$ is added to the charge, $e\rightarrow q_j$, the reduced mass, $\mu \rightarrow \mu_j$, and resonant frequency, $\omega^0 \rightarrow \omega_j^0$. Thus \eqref{eqn:17} is replaced by 
\begin{equation}\label{eqn:18}
\langle p^{V\!F}_j\!\rangle=\frac{(q_j^2/\mu_j)\mathbb{E}_0}{(\omega_j^0)^2} \, .
\end{equation}
Eq.  \eqref{eqn:18}  is the expectation value of an electric dipole that occurs in the vacuum as a result of  a VF being present, interacting with a photon, and becoming  polarized.  

The electric field of a photon that interacts with a parapositronium VF is the electric field of the photon at the instant $t_i$ that the photon becomes part of the polarized VF state.  The photon released in the decay of that state is identical to the incident photon. (See the ``Appendix''.)   The mathematical result is that the calculated expression for the permittivity $\epsilon_0$ of the vacuum does not depend on the frequency $\omega$ of the incident photon, implying that the behavior is not resonant.


After a parapositronium VF interacts with a photon, the photon-excited parapositronum VF is much more stable than the  original parapositronium VF. From  \eqref{eqn:8} the average lifetime of a parapositronium VF is $\Delta t_{p-Ps}\cong 3.2 \times10^{-22}$s. In contrast, from footnote 3  or  \eqref{eqn:57}, the mean lifetime of photon-excited parapositronum VF is  $1/\Gamma_{p-Ps}\cong 6.2 \times 10^{-11}$s.  The photon-excited parapositronium VF is much more stable than a parapositronium VF and is known as a quasi-stationary state \cite{Davydov:76}. 

The energy, momentum and angular momentum of the photon are all transferred to the parapositronium VF-photon combination during the interaction.  Such a state with an added photon has been has been described by Dodonov\cite{Dodonov:02}.  When the parapositronium atom vanishes back into the vacuum, a photon is emitted that is identical to the incident photon as described by the QED calculation in the ``Appendix''.

As \eqref{eqn:5} indicates, in the center of mass system the parapositronium VF can be described as a one-dimensional harmonic oscillator.  In that description, in addition to imposing the Heisenberg uncertainty principle of \eqref{eqn:7} as an equality, because the harmonic oscillator is one-dimensional there is a second Heisenberg uncertainly principle
\begin{equation}\label{eqn:1800}
\Delta x \Delta p  \geq \frac{\hbar}{2}   \,,
\end{equation}
to which the equality sign can be applied, producing a state of minimum uncertainty \cite{Gersch:92} for the wave packet $\psi (x)$ of the oscillator.  In particular, the one-dimensional wave packet will not change in character during the time for which the VF exists.  When the photon interacts with the VF,  a polarized, photon-excited VF is  created, delaying the progress of the photon until the  polarized, photon-excited VF decays.

Note that the polarization of bound, charged lepton-antilepton VFs by the presence of a real test photon, which has just been calculated in \eqref{eqn:17} for parapositronium, is different from vacuum polarization, a subject originally discussed by Serber\cite{Serber:35} and Uehling\cite{Uehling:35} and treated in many texts\cite{Gottfried:86,Bjorken:64,Greiner:94,Zee:10}.  Vacuum polarization results from the presence of virtual,  charged lepton-antilepton pairs created from either virtual or real photons and is calculated using perturbation theory in quantum electrodynamics.  Feynman diagrams for vacuum polarization are, for example, given in \cite{Zee:10}.  Although vacuum polarization is a separate process from the polarization of a VF, vacuum polarization must be taken into account when considering charged lepton-antilepton VFs, in that it is the charge renormalized by vacuum polarization that constitutes the charge that is indirectly observed. 

 \section{Calculation of dielectric properties of the vacuum: permittivity of the vacuum} 
\label{sec:5}

In a dielectric\cite{Jackson:99} the electric displacement $D(t)$ satisfies 
\begin{equation}\label{eqn:19}
D(t)=\epsilon E(t)=\epsilon_0 E(t) +P(t)\,,
\end{equation}
where
\begin{equation}\label{eqn:20}
P(t)=\sum_j N_j p_j(t)\,.
\end{equation}
In  \eqref{eqn:19} $\epsilon$ is the permittivity of the dielectric, and  $P(t)$ is the  polarization density.  In \eqref{eqn:20}   $N_j$ is the number of oscillators per unit volume of the $j^{\rm th}$ variety that  are available to interact, and $p_j(t)$ is the average dipole moment of the $j^{\rm th}$ variety of oscillator. 

The polarization density $P(t)$ is responsible for the increase from  $\epsilon_0 E(t)$ to $\epsilon E(t)$  because of photons interacting  with oscillators in the dielectric and results entirely from polarization of the atoms, molecules or both in  the dielectric. It then follows that in  the vacuum $\epsilon_0 E(t)$ must result entirely from the polarization density $P^{V\!F}(t)$ of atoms, molecules, or both that are VFs.  Thus, 
\begin{equation}\label{eqn:21}
\epsilon_0  = \frac{P^{V\!F}(t)}{E(t) }\,. 
\end{equation}
Using  \eqref{eqn:20} and then \eqref{eqn:18},  \eqref{eqn:21} becomes
\begin{equation}\label{eqn:22}
\epsilon_0 = \sum_j\frac{ N^{V\!F}_j \langle p_j^{V\!F}\!\rangle}{\mathbb{E}_0} = \sum_j N^{V\!F}_j \frac{(q_j^2/\mu_j)}{(\omega_j^0)^2}\,.
\end{equation}
The three types of VFs considered first are bound states of a charged lepton and antilepton, namely, parapositronium, muon-antimuon bound states, and tau-antitau bound states. Quark-antiquark states will  be discussed later.   Again, initially attention is restricted to parapositronium that is a VF. 

The progress of a photon traveling through the vacuum is slowed when it interacts with and has a polarizing effect on a VF consisting of a charged lepton and antilepton bound into an atom in its ground state.  Isolated ordinary matter is kinematically forbidden from absorbing and then reemitting a photon. But an isolated bound state of a charged particle-antiparticle VF can be polarized by a photon and then, when the polarized VF annihilates, emit an identical photon,
returning to the vacuum the energy originally borrowed for the creation of the VF.  From the ``Appendix'', after reinstating factors of $\hbar$ and $c$, the electromagnetic decay rate $\Gamma_{\rm p-Ps}$  for a VF parapositronium atom after it has interacted with the incident photon to form a  quasi-stationary  state is
\begin{equation}\label{eqn:23}
\Gamma_{\rm p-Ps} = \frac{\alpha^5 m_e c^2}{ \hbar} \, .
\end{equation}
The decay rate of ordinary parapositronium into two photons is half the above rate.\cite{Wheeler:46,Jauch:76}.  

The  quantity $e^{-\Gamma_{\rm p-Ps}t}$ is the probability that a photon-excited  parapositronium VF has not decayed electromagnetically during a time $t$, and $1-e^{-\Gamma_{\rm p-Ps}t}$ is the probability that it has  decayed.  The number density of  parapositronium VFs  with which a photon actually interacts is  $N_j^{VF}$ in  \eqref{eqn:22}.  The rate for a polarized parapositronium VF to  annihilate and emit a photon equals the rate for a parapositronium VF to  interact with a photon and form a polarized VF state.  As a result the quantity $1-e^{-\Gamma_{\rm p-Ps}t}$ is the probability that  a parapositronium VF absorbs a photon during a time $t$.

The quantity $N^{VF}_j$ for a parapositronium VF, denoted  $N_{\rm p-Ps}$,  equals the number density of parapositronium VFs  times the probability that a parapositronium VF   will interact with an incoming photon during the lifetime $\Delta t_{\rm p-Ps}$: 
\begin{align}\label{eqn:24}
N_{\rm p-Ps}&\cong \frac{1}{(L_{\rm p-Ps})^3}\times(1- e^{-\Gamma_{\rm p-Ps}\, \Delta t_{\rm p-Ps}}) \,.
\end{align}
Since $\Gamma_{\rm p-Ps}\,\Delta t_{\rm p-Ps}\ll1$,  $1- e^{-\Gamma_{\rm p-Ps}\, \Delta t_{\rm p-Ps}}$ can be replaced by $\Gamma_{\rm p-Ps}\,\Delta t_{\rm p-Ps}$, with the result 
\begin{align}\label{eqn:25}
N_{\rm p-Ps} \cong \frac{1}{(L_{\rm p-Ps})^3}\times \Gamma_{\rm p-Ps}\Delta t_{\rm p-Ps}=  \frac{\alpha^5}{4}{\left ({\frac{4 m_e c}{\hbar}} \right )}^3\,.
\end{align}

Substituting \eqref{eqn:4} and \eqref{eqn:25} into \eqref{eqn:22},
\begin{equation}\label{eqn:26}
\epsilon_0 \cong  \sum_j  \frac{ 8^3\alpha e^2}{\hbar c} \, .
\end{equation}
Note that the mass of the electron has cancelled from the expression for $\epsilon_0$, implying that bound muon-antimuon  and tau-antitau VFs each contribute the same amount to the value of  $\epsilon_0$ as parapositronium VFs. Including the  contributions from the three types of  bound, charged lepton-antilepton VFs yields
\begin{equation}\label{eqn:27}
\epsilon_0 \cong 3\,  \frac{ 8^3\alpha e^2}{\hbar c}\, +\substack{\mbox{possible quark-antiquark}\\ \\\mbox{ VF contributions.}}
\end{equation}

Quark-antiquark VFs contribute little to the value of $\epsilon_0$: for the heavy quarks $Q=c, b,\, {\rm or}\, t$ it is appropriate to think in terms of static quark potentials for $Q{\bar Q}$ bound states. An analysis similar to that leading to \eqref{eqn:26}  shows that the contribution to $\epsilon_0$ from $\eta_c(1S)$, the least massive $c{\bar c}$ bound state that has $J=0$ and positive charge conjugation parity\cite{Patrignani:16},  is at least $10^{-4}$ times smaller than the combined contribution from the three charged lepton-antilepton VFs. The corresponding $b{\bar b}$ and  $t{\bar t}$ bound states are estimated to contribute even less.  Consider the light quarks $q=u, d,\, {\rm and}\, s$: the $\pi^0, \eta$, and $\eta^\prime$  are the least massive $q{\bar q}$ bound states that have $J=0$ and decay into two photons.  Since the strong interactions are primarily responsible for the binding of these relativistic states,  $q{\bar q}$ bound state VFs would have much higher natural frequencies than the electromagnetically bound,  charged lepton-antilepton VFs.  Accordingly $q{\bar q}$ bound state VFs  contribute little to $\epsilon_0$ and to lowest order need not be considered.

Ignoring the small contributions to $\epsilon_0$ from quark-antiquark VFs, an approximate formula for  $\epsilon_0$  is immediately obtained from \eqref{eqn:27} using the defining formula $\alpha=e^2/(4 \pi \epsilon_0 \hbar c)$ to eliminate $\alpha$ and then using $c=1/\sqrt{\mu_0\epsilon_0}$: 
\begin{equation}\label{eqn:28}
\epsilon_0 \cong   \frac{ 6\mu_0}{\pi}\left(\frac{8e^2}{\hbar}\right)^2= 9.10\times 10^{-12}\rm \frac{C}{Vm}\, .
\end{equation}
Only the lowest-order terms in $\alpha$ have been retained in calculating $\epsilon_0$.  The  binding energy of parapositronium was neglected when calculating $\Delta t_{p-Ps}$ in  \eqref{eqn:8},  and   only the leading term was retained when calculating $\omega_j^0$. Moreover,  only the leading term in the formula for $\Gamma_{\rm p-Ps}$ has been included. 

\section{Permeability of the vacuum } 
\label{sec:6}

In \eqref{eqn:2} a formula for the speed of light $c$ was calculated using \eqref{eqn:1} and $c=1/\sqrt{\mu_0\epsilon_0}$, which is obtained from Maxwell's equations provided that the electric displacement $\mathbf{D}=\epsilon_0\mathbf{E}$, where $\mathbf{E}$ is the electric field; the magnetic field  $\mathbf{B}=\mu_0\mathbf{H}$, where $\mathbf{H}$ is the magnetic field intensity. However, in a medium\cite{Jackson:99}
\begin{equation}\label{eqn:1028}
\mathbf{B}=\mu_0(\mathbf{H}+\mathbf{M})\,,
\end{equation}
where $\mathbf{M}$ is the magnetic moment density of the medium. In the fermion quantum vacuum the medium consists of VFs. Parapositronium has no magnetic moment resulting from the orbital motion of the electron and positron because parapositronium is a singlet state, the electron and positron have equal masses and the charges on the electron and positron are equal in magnitude and opposite in sign.  

Also, for parapositronium the expectation value of the combined magnetic moments of the electron and positron is zero\cite{Sauder:67}. Thus for parapositronium $\mathbf{M}=0$.  (Similarly $\mathbf{M}=0$ for muon-antimuon, tau-antitau and quark-antiquark VFs.)  Thus, in spite of the presence of VFs, the vacuum satisfies the standard vacuum relation $\mathbf{B}=\mu_0\mathbf{H}$, justifying the  statement in footnote 3 that for lepton-antilepton VFs, $\mu=\mu_0$, and justifying the calculations of $\epsilon_0$ and $c$ in \eqref{eqn:1} and \eqref{eqn:2}, respectively.

\section{Results and Discussion} 
\label{sec:7}

In the model  of the the vacuum  discussed here, VFs of bound, charged lepton-antilepton pairs are polarized by photons much the way that ordinary matter is polarized. As a consequence,  the permittivity of the vacuum \eqref{eqn:28} and then $c$ \eqref{eqn:2} are calculated. The calculation of $c$ presented here automatically satisfies the condition that $c$ is the same in any direction in every inertial frame. This eliminates the need for one of the two postulates on which special relativity is based.

In the  early universe when the temperature was  sufficiently high that it was difficult for charged lepton-antilepton  VFs to either bind into atoms or remain bound once they had formed atoms,  the number density of  charged lepton-antilepton atoms that are VFs would have  been less than today. From \eqref{eqn:22} it then follows that $\epsilon_0$ would also have been smaller.   Since $c=1/\sqrt{\mu_0\epsilon_0}$,  a decrease in the value of $\epsilon_0$ would  increase the speed of light.  Variable-speed-of-light cosmology, for example, has been discussed, originally by Moffat\cite{Moffat:93} and later by Magueijo\cite{Magueijo:03}. The finite temperature and density corrections to vacuum polarization must be taken into account when describing the early universe\cite{Ahmed:91}. The results obtained here should provide additional insight into such investigations.

Various books have summarized physicists' significantly expanded understanding of processes involving photons. These include the books by Milonni\cite{Milonni:94},  Meis\cite{Meis:17a}, Leonhardt\cite{Leonhardt:97},  {\it Quantum Optics} by Garrison and Chiao\cite{Garrison:08}, and  {\it Quantum Optics} by Grynberg, Aspect and Fabre\cite{Grynberg:10}.    There is, however, much yet to be learned.  In particular, a further analysis is needed of fluctuations of the photon vacuum, carried out along the lines presented here for fermion-antifermion VFs.

\begin{appendix}
\section{Calculation of the decay rate of a photon-excited, parapositronium vacuum fluctuation}
\label{Sec:8}
\renewcommand{\theequation}{\Alph{section}.\arabic{equation}}
\setcounter{equation}{0}

Here the decay rate is calculated for a photon-excited, VF  of parapositronium. Single-photon decay is kinematically forbidden for a photon-excited, quasi-stationary state of ordinary parapositronium but is allowed for  a photon-excited, quasi-stationary state of parapositronium that is a VF since its  energy and momentum are transitory and thus do not enter into overall energy-momentum conservation.  The formula for the decay rate immediately generalizes to yield decay rates for VFs of  muon-antimuon and tau-antitau that are bound into atoms in their singlet, ground states.  
 
 Labeling the initial (incident) and final (emitted) photons, respectively, by $\gamma_i$ and $\gamma_f$, to lowest order the two Feynman diagrams that contribute to the process $\gamma_i$+positronium  that is a VF $\rightarrow \gamma_f$ are shown in Fig.~\ref{fig:1}.\footnote{Theorists' units are used in this ``Appendix'', implying that $\hbar$ and $c$ are each replaced by 1.}
\setcounter{figure}{0}
\begin{figure}[h]
\vspace{-0.3 cm}
\begin{center}
\includegraphics[width=90mm]{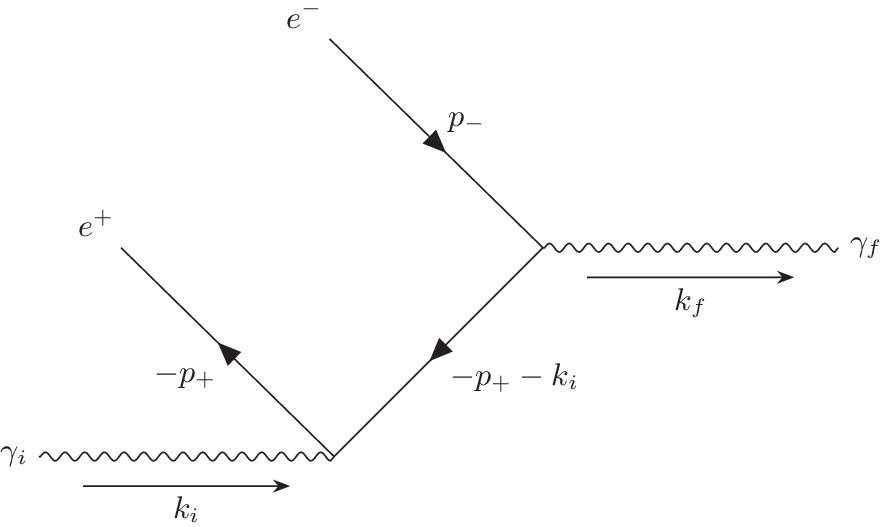}
\end{center}
\begin{center}
\vspace{-0.7 cm}(a)
\end{center}
\vspace{-0.7 cm}
\begin{center}
\includegraphics[width=70mm]{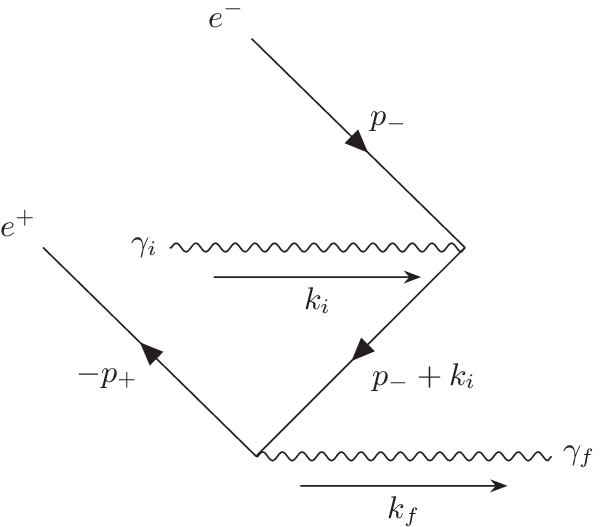}
\end{center}
\begin{center}
\vspace{-0.7 cm}(b)
\end{center}
\vspace{-0.5 cm}
\caption{ (a) Photon $\gamma_i$ interacts with a positron that then annihilates with an electron, emitting  photon $\gamma_f$. (b) Photon  $\gamma_i$ interacts with an electron that then annihilates with a positron, emitting  photon $\gamma_f$.}
\label{fig:1}
\end{figure}
In the  diagrams $p_-, p_+, k_i,$ and $k_f$ are, respectively, the four-momenta of the electron, positron, initial  photon, and final photon.

As already mentioned, for ordinary positronium the process is kinematically forbidden.  In the center-of-mass rest frame of positronium, $\mathbf{p_-}+\mathbf{p_+}=0$. Therefore, in this frame,
\begin{subequations}\label{eqn:32}
\begin{align}
\label{eqn:32a}
p_-&=(E_-, \mathbf{p_-})=(\sqrt{m_e^2+\mathbf{p_-}^2},\mathbf{p_-})\,, \\
\label{eqn:32b}
p_+&=(E_+, \mathbf{p_+})=(\sqrt{m_e^2+\mathbf{p_+}^2},\mathbf{p_+})\,,\nonumber\\
&=(\sqrt{m_e^2+\mathbf{p_-}^2},-\mathbf{p_-})=(E_-, -\mathbf{p_-})\,,\\
\label{eqn:32c}
k_i&=(\omega_i,\mathbf{k_i})=(|\mathbf{k_i}|,\mathbf{k_i})\,,  \\
\label{eqn:32d}
k_f&=(\omega_f,\mathbf{k_f})=(|\mathbf{k_f}|,\mathbf{k_f})\,, 
\end{align}
\end{subequations}
where $m_e$ is the mass of an electron. Conservation of energy and momentum requires $p_++p_-+k_i=k_f$.  Squaring both sides of the above equation yields
\begin{equation}\label{eqn:33}
E_-^2+E_-\omega_i=0\,.
\end{equation}
Eq.  \eqref{eqn:33} cannot be satisfied for ordinary positronium since both terms on the left-hand side are positive. However, after a photon excites   positronuim that is a VF, a photon can be emitted, but only when the positronium vanishes into the vacuum because only then does $E_-\rightarrow 0$, allowing  \eqref{eqn:33} to be satisfied.

When performing an electrodynamics calculation, if a factor exp$\pm (ip\cdot x)$ is associated with a particle that is part of a VF when it appears in its initial state, a factor  exp$\mp (ip\cdot x)$ is associated  with a particle that is part of a VF  when it vanishes into the vacuum.  This just eliminates the contribution of the particle that is part of a VF to overall energy-momentum conservation. When progressing along an energy-momentum line in a Feynman diagram, the energy-momentum associated with a particle that is part of a VF is not further used after the particle vanishes.  Since the  parapositronium atom is a VF, it is on shell. That is, as is the case for physical parapositronium, the electron and positron are on shell and they are bound by 6.8 eV, the binding energy of parapositronium.  Using the notation  of Ref.\cite{Bjorken:64}, the S-matrix for the transition photon +   positronium that is a  VF $\rightarrow$ photon is\footnote{The cross section is calculated for positronium that is a VF, and the restriction to parapositronium is not made until the decay rate is calculated.}
\begin{align}\label{eqn:34}
S_{\rm fi}=& \frac{e^2}{V^2}\sqrt{\frac{m_e}{E_+}}\sqrt{\frac{m_e}{E_ -}}\frac{1}{\sqrt{2\omega_i}}\frac{1}{\sqrt{2 \omega_f}}\,(2\pi)^4\delta(k_i-k_f) \times \nonumber \\
&\bar{v}(p_+,s_+)[(-i\slashed{\epsilon}_i)\frac{i}{-\slashed{p}_+ -\slashed{k}_i-m_e} (-i\slashed{\epsilon}_f)+\nonumber\\
&\hspace{1.5 cm}(-i\slashed{\epsilon}_f)\frac{i}{\slashed{p}_- +\slashed{k}_i-m_e} (-i\slashed{\epsilon}_i)]u(p_-,s_-)\,.
\end{align}
In \eqref{eqn:34} the fermion wave functions are normalized to unit probability in a box of volume $V$.  The equality $1/(\slashed{p}\pm m_e)=(\slashed{p}\mp m_e)/(p^2-m_e^2)$ is used to rewrite the above two propagators. From the identity $\{\gamma^\mu,\gamma^\nu\}=2g^{\mu\nu}I$, the equation
\begin{equation}\label{eqn:35}
\slashed{a} \slashed{b}=-\slashed{b} \slashed{a}+2a\cdot b I
\end{equation}
immediately follows where $a$ and $b$ are four-vectors and $I$ is the identity matrix.  Using  \eqref{eqn:35} $\slashed{\epsilon}_i \slashed{p}_+=-\slashed{p}_+ \slashed{\epsilon}_i+2p_+\cdot \epsilon_i I$ and 
$\slashed{p}_- \slashed{\epsilon}_i =-\slashed{\epsilon}_i\slashed{p}_-+2p_-\cdot \epsilon_i I$. Also $\bar{v}(p_+,s_+)\slashed{p}_+=-\bar{v}(p_+,s_+)\,m_e$ and $\slashed{p}_-u(p_-,s_-)=m_e\,u(p_-,s_-)$, with the result that \eqref{eqn:34} can be rewritten as
\begin{align}\label{eqn:36}
S_{\rm fi}=&-i \frac{e^2}{V^2}\sqrt{\frac{m_e}{E_+}}\sqrt{\frac{m_e}{E_ -}}\frac{1}{\sqrt{2\omega_i}}\frac{1}{\sqrt{2 \omega_f}}\,(2\pi)^4\delta(k_i-k_f) \times \nonumber \\
&\bar{v}(p_+,s_+)\left[-\frac{ (2\epsilon_i\cdot p_++\slashed{\epsilon}_i\slashed{k}_i) \slashed{\epsilon}_f}{(p_++k_i)^2-m_e^2}+\right.\nonumber\\ 
&\hspace{2.0 cm}\left. \frac{\slashed{\epsilon}_f(2p_-\cdot\epsilon_i+\slashed{k}_i\slashed{\epsilon}_i)}{(p_-+k_i)^2-m_e^2}\right]u(p_-,s_-)\,.
\end{align}

To obtain the decay rate to lowest order, in $S_{fi}$ the  respective velocities $\mathbf{v_-}$ and $\mathbf{v_+}$  of the electron and positron can be neglected. Thus 
\begin{subequations}\label{eqn:37}
\begin{align}\label{eqn:37a}
&E_- \rightarrow m_e\,, \hspace{1.0 cm} E_+ \rightarrow m_e\,,\\
\label{eqn:37b}
 &p_\pm \rightarrow (m_e, \mathbf{0})\, .
 \end{align}
\end{subequations}
The respective polarization vectors $\epsilon_i$ and $\epsilon_f$ of the initial and final photons are chosen to be space-like: $\epsilon_i=(0,\boldsymbol{\epsilon_i})$,  $\epsilon_f=(0,\boldsymbol{\epsilon_f})$ where
\begin{subequations}\label{eqn:38}
\begin{align}\label{eqn:38a}
&k_i \cdot \epsilon_i=-\mathbf{k_i}\cdot \boldsymbol{\epsilon_i}=0\,,\\
\label{eqn:38b}
&k_f\cdot \epsilon_f=-\mathbf{k_f}\cdot \boldsymbol{\epsilon_f}=0\,.
\end{align}
\end{subequations}
Using \eqref{eqn:37}, \eqref{eqn:38}, and $k_i\cdot k_i=0$,
\begin{subequations}\label{eqn:39}
\begin{align}\label{eqn:39a}
&(p_\pm+k_i)^2-m_e^2=2m_e\omega_i\,,\\
\label{eqn:39b}
&\epsilon_i \cdot p_\pm=0\,.
\end{align}
\end{subequations}
With the aid of \eqref{eqn:39}, \eqref{eqn:36} becomes
\begin{align}\label{eqn:40}
S_{\rm fi}=&-i \frac{e^2}{V^2}\frac{1}{\sqrt{2\omega_i}}\frac{1}{\sqrt{2 \omega_f}}\frac{1}{2m_e \omega_i}\,(2\pi)^4\delta^4(k_i-k_f) \times \nonumber \\
&\bar{v}(p_+,s_+)(- \slashed{\epsilon}_i\slashed{k}_i \slashed{\epsilon}_f
+\slashed{\epsilon}_f\slashed{k}_i\slashed{\epsilon}_i)u(p_-,s_-)\,.
\end{align}

Since $k_i=k_f$,  it follows from \eqref{eqn:35} and \eqref{eqn:38} that $\slashed{k_i}$ anti-commutes  with both $\slashed{\epsilon}_i$ and  $\slashed{\epsilon}_f$, allowing \eqref{eqn:40} to be rewritten as
\begin{align}\label{eqn:41}
S_{\rm fi}=&-i \frac{e^2}{V^2}\frac{1}{\sqrt{2\omega_i}}\frac{1}{\sqrt{2 \omega_f}}\frac{1}{2m_e \omega_i}\,(2\pi)^4\delta^4(k_i-k_f) \times \nonumber \\
&\bar{v}(p_+,s_+)(\slashed{\epsilon}_i \slashed{\epsilon}_f
-\slashed{\epsilon}_f\slashed{\epsilon}_i)\slashed{k}_iu(p_-,s_-)\,.
\end{align}
Then,
\begin{align}\label{eqn:42}
|S_{\rm fi}|^2=&\frac{e^4}{V^4}\frac{1}{2\omega_i}\frac{1}{2 \omega_f}\frac{1}{4m_e^2 \omega_i^2}VT\,(2\pi)^4\delta^4(k_i-k_f) \times \nonumber \\
&\bar{v}(p_+,s_+)(\slashed{\epsilon}_i \slashed{\epsilon}_f
-\slashed{\epsilon}_f\slashed{\epsilon}_i)\slashed{k}_i u(p_-,s_-)\times \nonumber\\
&\bar{u}(p_-,s_-)\slashed{k}_i(\slashed{\epsilon}_f \slashed{\epsilon}_i
-\slashed{\epsilon}_i\slashed{\epsilon}_f) v(p_+,s_+)\,.
\end{align}
In \eqref{eqn:42} the interaction is assumed to occur in the time interval $-T/2<t<T/2$.  

Summing over the electron and positron spins, which converts $|S_{\rm fi}|^2$ into a trace denoted by $Tr$,
\begin{align}\label{eqn:43}
&\sum_{s_\pm} |S_{\rm fi}|^2=\frac{e^4}{V^4}\frac{1}{2\omega_i}\frac{1}{2 \omega_f}\frac{1}{16m_e^4 \omega_i^2}VT\,(2\pi)^4\delta^4(k_i-k_f)\times \nonumber \\ 
&Tr[(\slashed{p}_+-m_e)(\slashed{\epsilon}_i \slashed{\epsilon}_f-\slashed{\epsilon}_f\slashed{\epsilon}_i) 
\slashed{k}_i (\slashed{p}_-+m_e)\slashed{k}_i(\slashed{\epsilon}_f \slashed{\epsilon}_i
-\slashed{\epsilon}_i\slashed{\epsilon}_f)]\,.
\end{align}
Using \eqref{eqn:35} to reverse the order of $\slashed{k}_i\slashed{p}_-$ and  using $\slashed{k}_i\slashed{k}_i=k_i\cdot k_i\,I=0$, the following term that appears in the second line above can be simplified:
\begin{equation}\label{eqn:44}
\slashed{k}_i (\slashed{p}_-+m_e)\slashed{k}_i=2m_e\omega_i\slashed{k}_i\,.
\end{equation}
As mentioned previously, $\slashed{k}_i$ anti-commutes with both $\slashed{\epsilon}_i$ and  $\slashed{\epsilon}_f$, with the result that \eqref{eqn:43} becomes
\begin{align}\label{eqn:45}
&\sum_{s_\pm}  |S_{\rm fi}|^2=\frac{e^4}{V^4}\frac{1}{2\omega_i}\frac{1}{2 \omega_f}\frac{1}{8m_e^3 \omega_i}VT\,(2\pi)^4\delta^4(k_i-k_f) \times \nonumber \\
&Tr[(\slashed{p}_+-m_e)(\slashed{\epsilon}_i \slashed{\epsilon}_i-\slashed{\epsilon}_f\slashed{\epsilon}_i)(\slashed{\epsilon}_f \slashed{\epsilon}_i
-\slashed{\epsilon}_i\slashed{\epsilon}_f)\slashed{k}_i]\,.
\end{align}
Eq. \eqref{eqn:35} is used as necessary to reverse the order of $\slashed{\epsilon}_i$ and $\slashed{\epsilon}_f$ so as to obtain terms of the form $\slashed{\epsilon}_i\slashed{\epsilon}_i=\epsilon_i \cdot \epsilon_i\,I=- \boldsymbol{\epsilon}_i \cdot \boldsymbol{\epsilon}_i\,I=-I$ and $\slashed{\epsilon}_f\slashed{\epsilon}_f=-I.$ All nonzero traces in \eqref{eqn:45}  are then either of the form $Tr(\slashed{a}\slashed{b})$ or $Tr(\slashed{a}\slashed{b}\slashed{c}\slashed{d}$) that are easily simplified, yielding the result
\begin{align}\label{eqn:46}
\sum_{s_\pm}  |S_{\rm fi}|^2=&\frac{2}{m_e^2}\frac{e^4}{V^4}\frac{1}{2\omega_i}\frac{1}{2 \omega_f}VT\,(2\pi)^4\delta^4(k_i-k_f) \times \nonumber \\
&[1-(\epsilon_i \cdot \epsilon_f)^2]\,.
\end{align}

The average cross section is now calculated for photon-excited positronium that is a VF to annihilate and emit a photon: $|S_{fi}|^2$ is divided by $VT$ to form a rate per unit volume, divided by the  electron-positron flux $|\mathbf{v}_+-\mathbf{v}_-|/V$, and divided by the number of target particles per unit volume $1/V$.  Averaging over the spins of the electron and positron ($\frac{1}{4}\sum_{s_\pm})$,  summing over the polarizations of the final photon ($\sum_{\epsilon_f})$, averaging over the polarizations of the initial photon ($\frac{1}{2}\sum_{\epsilon_i})$, summing over the number of states  of the final photon in the momentum interval ${\rm d}^3\mathbf{k}_f$ $(\int V {\rm d}^3\mathbf{k}_f/(2\pi)^3)$, and averaging over the number of states  of the initial photon in the momentum interval ${\rm d}^3\mathbf{k}_i$ $(\int V {\rm d}^3\mathbf{k}_i/(2\pi)^3)$, an expression for the cross section $\sigma$ is obtained:
\begin{align}\label{eqn:47}
\sigma=&\frac{1}{4}\sum_{s_\pm}\sum_{\epsilon_f}\frac{1}{2}\sum_{\epsilon_i}\int \frac{V{\rm d}^3\mathbf{k}_f}{(2\pi)^3} \int \frac{V{\rm d}^3\mathbf{k}_i}{(2\pi)^3}\times\nonumber\\
&\frac{|S_{fi}|^2}{VT}\frac{1}{\frac{|\mathbf{v}_+-\mathbf{v}_-|}{V}}\frac{1}{\frac{1}{V}} \,.
\end{align}
Using \eqref{eqn:46}, \eqref{eqn:47} becomes
\begin{align}\label{eqn:48}
\sigma=&\frac{\alpha^2}{m_e^2}\frac{1}{|\mathbf{v}_+-\mathbf{v}_-|}\sum_{\epsilon_f}\sum_{\epsilon_i}[1-(\epsilon_i \cdot \epsilon_f)^2]\times\nonumber\\
&\int_{-\infty}^\infty \frac{{\rm d}^3\mathbf{k}_f}{2\omega_f} \int_{-\infty}^\infty \frac{{\rm d}^3\mathbf{k}_i}{2\omega_i} \delta^4(k_i-k_f) \,.
\end{align}
Choosing the z-axis to point in the direction of $\mathbf{k_i}$, the unit polarization vectors for the initial photon  $\boldsymbol{\epsilon}_i^a$ and  $\boldsymbol{\epsilon}_i^b$ are chosen in the x- and y-direction, respectively.  Because the delta function in  \eqref{eqn:48} imposes the condition $k_f=k_i$, the unit polarization vectors for the final photon $\boldsymbol{\epsilon}_f^a$ and  $\boldsymbol{\epsilon}_f^b$ can also be chosen in the x- and y-direction, respectively. The sum over polarizations in \eqref{eqn:47} is  now easily performed:
\begin{subequations}\label{eqn:49}
\begin{align} \label{eqn:49a}
&\sum_{\epsilon_f}\sum_{\epsilon_i} 1=4\,,\\
\label{eqn:49b}
&\sum_{\epsilon_f}\sum_{\epsilon_i}(\epsilon_i \cdot \epsilon_f)^2=\nonumber\\
&(\epsilon_i^a \cdot \epsilon_f^a)^2+(\epsilon_i^b \cdot \epsilon_f^a)^2+(\epsilon_i^a \cdot \epsilon_f^b)^2+
(\epsilon_i^b \cdot \epsilon_f^b)^2\nonumber=\\
&(-1)^2+0+0+(-1)^2=2\,.
\end{align}
\end{subequations}

Using \eqref{eqn:32d} and the identity\cite{Schiff:55},
\begin{equation}\label{eqn:50}
\delta(\omega^2-a^2)=\frac{1}{2a}[\delta(\omega-a)+\delta(\omega+a)],\;a>0\,, 
\end{equation}
it is straightforward to show that
\begin{equation}\label{eqn:51}
\int_{-\infty}^\infty \frac{{\rm d}^3\mathbf{k}_i}{2\omega_i}=\int_{-\infty}^\infty {\rm d}^4k_i\,\delta(k_i^2) \theta(k_{i0})\,.
\end{equation}
The theta function $\theta(k_{i0})=0$ if $k_{i0}\langle0$ and $\theta(k_{i0})=1$ if $k_{i0}\rangle0$. With the aid of \eqref{eqn:51}, the second line in \eqref{eqn:48}  can be rewritten as
\begin{align}\label{eqn:52}
&\int_{-\infty}^\infty \frac{{\rm d}^3\mathbf{k}_f}{2\omega_f} \int_{-\infty}^\infty \frac{{\rm d}^3\mathbf{k}_i}{2\omega_i}\, \delta^4(k_i-k_f)=\nonumber\\
&\int_{-\infty}^\infty \frac{{\rm d}^3\mathbf{k}_f}{2\omega_f}\,\delta(k_f^2)\,\theta(k_{f0})\,.
\end{align}
Factoring $k_f^2=k_{f0}^2-|\mathbf{k}_f|^2$ in the above $\delta$-function, using \eqref{eqn:50}, rewriting  ${\rm d}^3\mathbf{k}_f$ as ${\rm d}\Omega_f
|\mathbf{k}_f|^2{\rm d}|\mathbf{k}_f|$, performing the angular integration over ${\rm d}\Omega_f$, which yields a factor of $4\pi$, and then integrating over $|\mathbf{k}_f|$,
\begin{equation}\label{eqn:53}
\int_{-\infty}^\infty \frac{{\rm d}^3\mathbf{k}_f}{2\omega_f} \int_{-\infty}^\infty \frac{{\rm d}^3\mathbf{k}_i}{2\omega_i}\, \delta^4(k_i-k_f)=\pi\,.
\end{equation}
Substituting \eqref{eqn:49} and \eqref{eqn:53} into \eqref{eqn:48} yields the formula for the cross section for the annihilation into a photon of photon-excited positronium that is a VF,
\begin{equation}\label{eqn:54}
\sigma=\frac{2 \pi \alpha^2}{m_e^2}\frac{1}{|\mathbf{v}_+-\mathbf{v}_-|} \,.
\end{equation}

From the formula for the cross section, a formula for the decay rate is readily obtained. The logic is the same as that used to calculate the decay rate  for parapositronium decaying into two photons\cite{Wheeler:46,Jauch:76}:  parapostronium, orthopositronium, and a photon have respective charge conjugation parities of +1, -1, and -1.  Thus photon-excited parapositronium has charge conjugation parity of -1 while photon-excited orthopositronium has charge conjugation parity of +1. Since electromagnetic interactions are invariant under charge conjugation,  photon-excited parapositronium, but not photon-excited orthopositronium, can decay into a single photon. 

In obtaining \eqref{eqn:54} the electron and positron spins were averaged over all four spins, resulting in the sum being divided by four. But the annihilating state is parapositronium, the singlet state.  Orthopositronium, the triplet state, does not contribute.  Since only one of the four spin states contributes to the cross section, the formula for the cross section should not have been divided by four, it should have been divided by the number one.  Thus the formula for $\sigma$ in \eqref{eqn:54} should be multiplied by a factor of four to obtain the cross section, abbreviated $\sigma_{\rm p-Ps}$, for the annihilation into a photon of  photon-excited parapositronium that is a VF,
\begin{equation}\label{eqn:55}
\sigma_{\rm p-Ps}=\frac{8 \pi \alpha^2}{m_e^2}\frac{1}{|\mathbf{v}_+-\mathbf{v}_-|} \,.
\end{equation}

For the annihilation of photon-excited parapositronium that is a VF into a photon, the electromagnetic  decay  rate $\Gamma_{p-Ps}$ is calculated using the mechanism for the annihilation of ordinary parapositronium\cite{Wheeler:46,Jauch:76}. The Schr\"odinger wave function $\psi(x)$ for parapositronium is just the ground-state hydrogen atom wave function with the reduced mass of hydrogen, which is approximately $m_e$, replaced by  $m_e/2$, the reduced mass of parapositronium:
\begin{equation}\label{eqn:56}
\psi(x)=\frac{1}{\sqrt{\pi}}\left(\frac{\alpha m_e}{2}\right)^{3/2}e^{- \alpha m_e\,r/2}\,.
\end{equation}
In the above formula $x$ is the magnitude of the vector $\mathbf{x}=\mathbf{x}_e-\mathbf{x}_p$ where $\mathbf{x}_e$ and $\mathbf{x}_p$ are, respectively,  the positions of the electron and the positron.  

The decay rate $\Gamma_{p-Ps}$ is the product of $\sigma_{\rm p-Ps}$ and the flux of a parapositronium atom, which is the relative velocity of approach of the electron and positron in parapositronium multiplied by $|\psi(0)|^2$, the probability density that the electron and positron collide and annihilate. 
\begin{align}\label{eqn:57}
\Gamma_{p-Ps}&=\sigma_{\rm p-Ps}|\mathbf{v}_+-\mathbf{v}_-|\,|\psi(0)|^2\,, \nonumber\\
&=\frac{8 \pi \alpha^2}{m_e^2}\frac{1}{|\mathbf{v}_+-\mathbf{v}_-|}|\mathbf{v}_+-\mathbf{v}_-|\,\frac{1}{\pi}\left(\frac{\alpha\,m_e}{2}\right)^3\,,\nonumber \\
&=\alpha^5m_e\,.
\end{align}
It immediately follows that the corresponding decay rates for  photon-excited,  muon-antimuon or tau-antitau  VFs bound into a singlet, ground state are obtained by replacing  the  electron mass in \eqref{eqn:57} with the muon or tau mass, respectively.  As photons travel through the vacuum,  these three decay rates characterize how photons interact with charged lepton-antilepton VFs.
\end{appendix}

\bibliography{Epsilon}
\end{document}